\pdfoutput=1
\documentclass[12pt,a4paper]{article}
\usepackage{ifthen} 
\newboolean{pdflatex}
\setboolean{pdflatex}{true} 

\newboolean{articletitles}
\setboolean{articletitles}{true} 

\newboolean{uprightparticles}
\setboolean{uprightparticles}{false} 

\newboolean{inbibliography}
\setboolean{inbibliography}{false} 

\usepackage[top=1in, bottom=1.25in, left=1in, right=1in]{geometry}

\usepackage{microtype}
\usepackage{lineno}  
\usepackage{xspace} 
\usepackage{caption}

\usepackage{graphicx}  
\usepackage{color}
\usepackage{colortbl}

\usepackage{amsmath} 
\usepackage{amssymb}
\usepackage{amsfonts}
\usepackage{upgreek} 

\newcommand*\patchAmsMathEnvironmentForLineno[1]{%
\expandafter\let\csname old#1\expandafter\endcsname\csname #1\endcsname
\expandafter\let\csname oldend#1\expandafter\endcsname\csname
end#1\endcsname
 \renewenvironment{#1}%
   {\linenomath\csname old#1\endcsname}%
   {\csname oldend#1\endcsname\endlinenomath}%
}
\newcommand*\patchBothAmsMathEnvironmentsForLineno[1]{%
  \patchAmsMathEnvironmentForLineno{#1}%
  \patchAmsMathEnvironmentForLineno{#1*}%
}
\AtBeginDocument{%
\patchBothAmsMathEnvironmentsForLineno{equation}%
\patchBothAmsMathEnvironmentsForLineno{align}%
\patchBothAmsMathEnvironmentsForLineno{flalign}%
\patchBothAmsMathEnvironmentsForLineno{alignat}%
\patchBothAmsMathEnvironmentsForLineno{gather}%
\patchBothAmsMathEnvironmentsForLineno{multline}%
\patchBothAmsMathEnvironmentsForLineno{eqnarray}%
}

\usepackage{hyperref}    
\usepackage[all]{hypcap}

\usepackage{xspace} 
\usepackage{upgreek}

\def\lhcb {\mbox{LHCb}\xspace}

\def\velo   {VELO\xspace}
\def\rich   {RICH\xspace}

\def\MagUp {\mbox{\em Mag\kern -0.05em Up}\xspace}

\ifthenelse{\boolean{uprightparticles}}%
{

 \def\Pmu         {\ensuremath{\upmu}\xspace}

 \def\Ppi         {\ensuremath{\uppi}\xspace}

 \def\Ppsi        {\ensuremath{\uppsi}\xspace}

 \def\PDelta      {\ensuremath{\Delta}\xspace}                 
 \def\PXi      {\ensuremath{\Xi}\xspace}                 
 \def\PLambda      {\ensuremath{\Lambda}\xspace}                 
 \def\PSigma      {\ensuremath{\Sigma}\xspace}                 
 \def\POmega      {\ensuremath{\Omega}\xspace}                 
 \def\PUpsilon      {\ensuremath{\Upsilon}\xspace}

 \def\PB      {\ensuremath{\mathrm{B}}\xspace}                 
                  
 \def\PD      {\ensuremath{\mathrm{D}}\xspace}

 \def\PJ      {\ensuremath{\mathrm{J}}\xspace}                 
 \def\PK      {\ensuremath{\mathrm{K}}\xspace}

 \def\Pb      {\ensuremath{\mathrm{b}}\xspace}                 
 \def\Pc      {\ensuremath{\mathrm{c}}\xspace}

 \def\Pi      {\ensuremath{\mathrm{i}}\xspace}

 \def\Pp      {\ensuremath{\mathrm{p}}\xspace}

}
{

 \def\Pmu         {\ensuremath{\mu}\xspace}

 \def\Ppi         {\ensuremath{\pi}\xspace}

 \def\Ppsi        {\ensuremath{\psi}\xspace}                 
                  
 \mathchardef\PDelta="7101
 \mathchardef\PXi="7104
 \mathchardef\PLambda="7103
 \mathchardef\PSigma="7106
 \mathchardef\POmega="710A
 \mathchardef\PUpsilon="7107
                  
 \def\PB      {\ensuremath{B}\xspace}                 
                  
 \def\PD      {\ensuremath{D}\xspace}

 \def\PJ      {\ensuremath{J}\xspace}                 
 \def\PK      {\ensuremath{K}\xspace}

 \def\Pb      {\ensuremath{b}\xspace}                 
 \def\Pc      {\ensuremath{c}\xspace}

 \def\Pi      {\ensuremath{i}\xspace}

 \def\Pp      {\ensuremath{p}\xspace}

}

\makeatletter
\ifcase \@ptsize \relax
  \newcommand{\miniscule}{\@setfontsize\miniscule{4}{5}}
\or
  \newcommand{\miniscule}{\@setfontsize\miniscule{5}{6}}
\or
  \newcommand{\miniscule}{\@setfontsize\miniscule{5}{6}}
\fi
\makeatother

\DeclareRobustCommand{\optbar}[1]{\shortstack{{\miniscule (\rule[.5ex]{1.25em}{.18mm})}
  \\ [-.7ex] $#1$}}

\def\mumu       {{\ensuremath{\Pmu^+\Pmu^-}}\xspace}

\def\cquark    {{\ensuremath{\Pc}}\xspace}

\def\bquark    {{\ensuremath{\Pb}}\xspace}

\def\pion   {{\ensuremath{\Ppi}}\xspace}

\def\pip    {{\ensuremath{\pion^+}}\xspace}
\def\pim    {{\ensuremath{\pion^-}}\xspace}

\def\kaon    {{\ensuremath{\PK}}\xspace}
  \def\Kbar    {{\kern 0.2em\overline{\kern -0.2em \PK}{}}\xspace}

\def\KorKbar    {\kern 0.18em\optbar{\kern -0.18em K}{}\xspace}

\def\Km      {{\ensuremath{\kaon^-}}\xspace}

\def\KS      {{\ensuremath{\kaon^0_{\mathrm{ \scriptscriptstyle S}}}}\xspace}

  \def\Dbar    {{\kern 0.2em\overline{\kern -0.2em \PD}{}}\xspace}
\def\D       {{\ensuremath{\PD}}\xspace}

\def\DorDbar    {\kern 0.18em\optbar{\kern -0.18em D}{}\xspace}
\def\Dz      {{\ensuremath{\D^0}}\xspace}

\def\Dstarp  {{\ensuremath{\D^{*+}}}\xspace}

\def\Bbar    {{\ensuremath{\kern 0.18em\overline{\kern -0.18em \PB}{}}}\xspace}

\def\BorBbar    {\kern 0.18em\optbar{\kern -0.18em B}{}\xspace}

\def\jpsi     {{\ensuremath{{\PJ\mskip -3mu/\mskip -2mu\Ppsi\mskip 2mu}}}\xspace}

  \def\Y#1S{\ensuremath{\PUpsilon{(#1S)}}\xspace}

\def\proton      {{\ensuremath{\Pp}}\xspace}

\def\Xires       {{\ensuremath{\PXi}}\xspace}

\def\Lz          {{\ensuremath{\PLambda}}\xspace}
\def\Lbar        {{\ensuremath{\kern 0.1em\overline{\kern -0.1em\PLambda}}}\xspace}
\def\LorLbar    {\kern 0.18em\optbar{\kern -0.18em \PLambda}{}\xspace}

\def\Lb      {{\ensuremath{\Lz^0_\bquark}}\xspace}

\def\Lc      {{\ensuremath{\Lz^+_\cquark}}\xspace}

\def\Xib     {{\ensuremath{\Xires_\bquark}}\xspace}

\def\Xibm    {{\ensuremath{\Xires^-_\bquark}}\xspace}

\def\Xicz    {{\ensuremath{\Xires^0_\cquark}}\xspace}

\def\BF         {{\ensuremath{\mathcal{B}}}\xspace}

\def\BR         {\BF}
\newcommand{\decay}[2]{\ensuremath{#1\!\to #2}\xspace}         

\def\to                 {\ensuremath{\rightarrow}\xspace}

\def\AT#1     {\ensuremath{A_{\mathrm{T}}^{#1}}\xspace}           

\def\C#1      {\ensuremath{\mathcal{C}_{#1}}\xspace}                       
\def\Cp#1     {\ensuremath{\mathcal{C}_{#1}^{'}}\xspace}                    
\def\Ceff#1   {\ensuremath{\mathcal{C}_{#1}^{\mathrm{(eff)}}}\xspace}        
\def\Cpeff#1  {\ensuremath{\mathcal{C}_{#1}^{'\mathrm{(eff)}}}\xspace}       
\def\Ope#1    {\ensuremath{\mathcal{O}_{#1}}\xspace}                       
\def\Opep#1   {\ensuremath{\mathcal{O}_{#1}^{'}}\xspace}                    

\newcommand{\tev}{\ifthenelse{\boolean{inbibliography}}{\ensuremath{~T\kern -0.05em eV}}{\ensuremath{\mathrm{\,Te\kern -0.1em V}}}\xspace}
\newcommand{\gev}{\ensuremath{\mathrm{\,Ge\kern -0.1em V}}\xspace}
\newcommand{\mev}{\ensuremath{\mathrm{\,Me\kern -0.1em V}}\xspace}
\newcommand{\kev}{\ensuremath{\mathrm{\,ke\kern -0.1em V}}\xspace}
\newcommand{\ev}{\ensuremath{\mathrm{\,e\kern -0.1em V}}\xspace}
\newcommand{\gevc}{\ensuremath{{\mathrm{\,Ge\kern -0.1em V\!/}c}}\xspace}
\newcommand{\mevc}{\ensuremath{{\mathrm{\,Me\kern -0.1em V\!/}c}}\xspace}
\newcommand{\gevcc}{\ensuremath{{\mathrm{\,Ge\kern -0.1em V\!/}c^2}}\xspace}
\newcommand{\gevgevcccc}{\ensuremath{{\mathrm{\,Ge\kern -0.1em V^2\!/}c^4}}\xspace}
\newcommand{\mevcc}{\ensuremath{{\mathrm{\,Me\kern -0.1em V\!/}c^2}}\xspace}

\def\mm   {\ensuremath{\mathrm{ \,mm}}\xspace}

\def\mum  {\ensuremath{{\,\upmu\mathrm{m}}}\xspace}

\def\invfb   {\ensuremath{\mbox{\,fb}^{-1}}\xspace}

\def\ps   {\ensuremath{{\mathrm{ \,ps}}}\xspace}

\newcommand{\stat}{\ensuremath{\mathrm{\,(stat)}}\xspace}
\newcommand{\syst}{\ensuremath{\mathrm{\,(syst)}}\xspace}

\newcommand{\chisq}{\ensuremath{\chi^2}\xspace}

\newcommand{\chisqip}{\ensuremath{\chi^2_{\text{IP}}}\xspace}

\def\gsim{{~\raise.15em\hbox{$>$}\kern-.85em
          \lower.35em\hbox{$\sim$}~}\xspace}
\def\lsim{{~\raise.15em\hbox{$<$}\kern-.85em
          \lower.35em\hbox{$\sim$}~}\xspace}

\def\sPlot{\mbox{\em sPlot}\xspace}

\def\ptot       {\mbox{$p$}\xspace}
\def\pt         {\mbox{$p_{\mathrm{ T}}$}\xspace}

\def\evtgen     {\mbox{\textsc{EvtGen}}\xspace}

\def\geant      {\mbox{\textsc{Geant4}}\xspace}

\def\photos     {\mbox{\textsc{Photos}}\xspace}

\def\pythia     {\mbox{\textsc{Pythia}}\xspace}

\def\tell1  {TELL1\xspace}
\def\ukl1   {UKL1\xspace}

\newcommand{\LbJpsipK}{\ensuremath{\Lb\to\jpsi\proton\Km}\xspace}

\newcommand{\LbJpsiL}{\ensuremath{\Lb\to\jpsi\Lz}\xspace}
\newcommand{\XbJpsiLK}{\ensuremath{\Xibm\to\jpsi\Lz\Km}\xspace}
\newcommand{\XbJpsiSK}{\ensuremath{\Xibm\to\jpsi\PSigma^{0}\Km}\xspace}
\newcommand{\SLg}{\ensuremath{\PSigma^{0}\to \Lz\gamma}\xspace}

\usepackage{cite} 
\usepackage{mciteplus}

\begin{document}

\renewcommand{\thefootnote}{\fnsymbol{footnote}}
\setcounter{footnote}{1}

\begin{titlepage}
\pagenumbering{roman}

\vspace*{-1.5cm}
\centerline{\large EUROPEAN ORGANIZATION FOR NUCLEAR RESEARCH (CERN)}
\vspace*{1.5cm}
\noindent
\begin{tabular*}{\linewidth}{lc@{\extracolsep{\fill}}r@{\extracolsep{0pt}}}
\ifthenelse{\boolean{pdflatex}}
{\vspace*{-2.7cm}\mbox{\!\!\!\includegraphics[width=.14\textwidth]{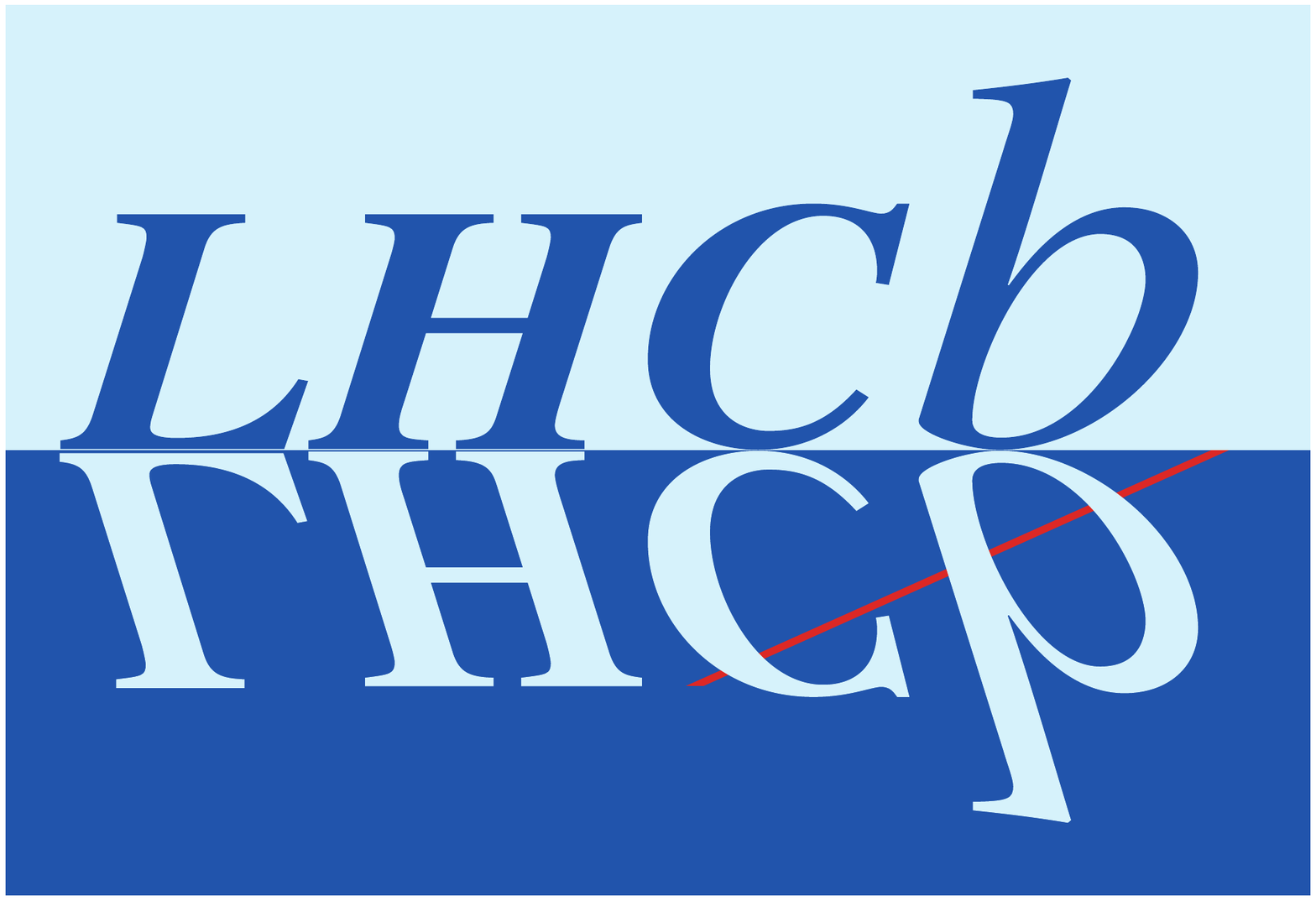}} & &}
{\vspace*{-1.2cm}\mbox{\!\!\!\includegraphics[width=.12\textwidth]{lhcb-logo.eps}} & &}
\\
 & & CERN-EP-2016-318 \\  
 & & LHCb-PAPER-2016-053 \\  
 & & 19 January 2017\\
\end{tabular*}

\vspace*{2.0cm}

{\normalfont\bfseries\boldmath\huge
\begin{center}
   Observation of the $\varXi_{b}^{-}\to J/\psi\varLambda K^{-}$ decay
\end{center}
}

\vspace*{2.0cm}

\begin{center}
The LHCb collaboration\footnote{Authors are listed at the end of this Letter.}
\end{center}

\vspace{\fill}

\begin{abstract}
The observation of the decay $\varXi_{b}^{-}\to J/\psi\varLambda K^{-}$ is reported, using a data sample corresponding to an integrated luminosity of $3~\mathrm{fb}^{-1}$, collected by the LHCb detector in $pp$ collisions at centre-of-mass energies of $7$ and $8~\mathrm{TeV}$.
The production rate of $\varXi_{b}^{-}$ baryons detected in the decay $\varXi_{b}^{-}\to J/\psi\varLambda K^{-}$ is measured relative to that of $\varLambda_{b}^{0}$ baryons using the decay  $\varLambda_{b}^{0}\to J/\psi \varLambda$. Integrated over the $b$-baryon transverse momentum $p_{\rm T}<25~\mathrm{GeV/}c $ and rapidity $2.0<y<4.5$, the measured ratio is
\begin{equation*}
\frac{f_{\varXi_{b}^{-}}}{f_{\varLambda_{b}^{0}}}\frac{\mathcal{B}(\varXi_{b}^{-}\to J/\psi\varLambda K^{-})}{\mathcal{B}(\varLambda_{b}^{0}\to J/\psi \varLambda)}=(4.19\pm 0.29~(\mathrm{stat})\pm0.15~(\mathrm{syst}))\times 10^{-2},
\end{equation*}where $f_{\varXi_{b}^{-}}$ and $f_{\varLambda_{b}^{0}}$ are the fragmentation fractions of $b\to\varXi_{b}^{-}$ and $b\to\varLambda_{b}^{0}$ transitions, and $\mathcal{B}$ represents the branching fraction of the corresponding $b$-baryon decay. The mass difference between $\varXi_{b}^{-}$ and $\varLambda_{b}^{0}$ baryons is measured to be 
\begin{equation*}
M(\varXi_{b}^{-})-M(\varLambda_{b}^{0})=177.08\pm0.47~(\mathrm{stat})\pm0.16~(\mathrm{syst} )~\mathrm{MeV/}c^{2}. 
\end{equation*}
\end{abstract}

\vspace*{1.0cm}

\begin{center}
Published in Phys.~Lett.~B 772 (2017) 265-273 
\end{center}

\vspace{\fill}

{\footnotesize 
\centerline{\copyright~CERN on behalf of the \lhcb collaboration, licence \href{http://creativecommons.org/licenses/by/4.0/}{CC-BY-4.0}.}}
\vspace*{2mm}

\end{titlepage}

\newpage
\setcounter{page}{2}
\mbox{~}
\cleardoublepage

\renewcommand{\thefootnote}{\arabic{footnote}}
\setcounter{footnote}{0}

\pagestyle{plain} 
\setcounter{page}{1}
\pagenumbering{arabic}

\section{Introduction}
\label{sec:introduction}

Since the birth of the quark model, the possibility of forming baryonic states from combinations of quarks other than three valence quarks has been considered~\cite{GellMann:1964nj,Zweig:1964}. For example, states with four quarks and an antiquark, referred to as pentaquarks~\cite{Lipkin:1987sk}, have been searched for experimentally for many years. 
As observed with the \lhcb detector at the LHC, the distribution of invariant mass of the $\jpsi p$ system in $\Lb\to\jpsi (\to\mu^+\mu^-)p K^-$ decays shows a narrow peak suggestive of $uudc\bar c$ pentaquark formation~\cite{LHCb-PAPER-2015-029,LHCb-PAPER-2015-032,LHCb-PAPER-2016-009}.  (The inclusion of charge conjugate processes is implied throughout the text.) From a six-dimensional amplitude model fit, two pentaquark resonances, decaying into $\jpsi p$, are observed with large significances~\cite{LHCb-PAPER-2015-029}. 

\begin{figure}[b]
\centering
\vskip -0.5cm
\includegraphics[width=1.0\textwidth]{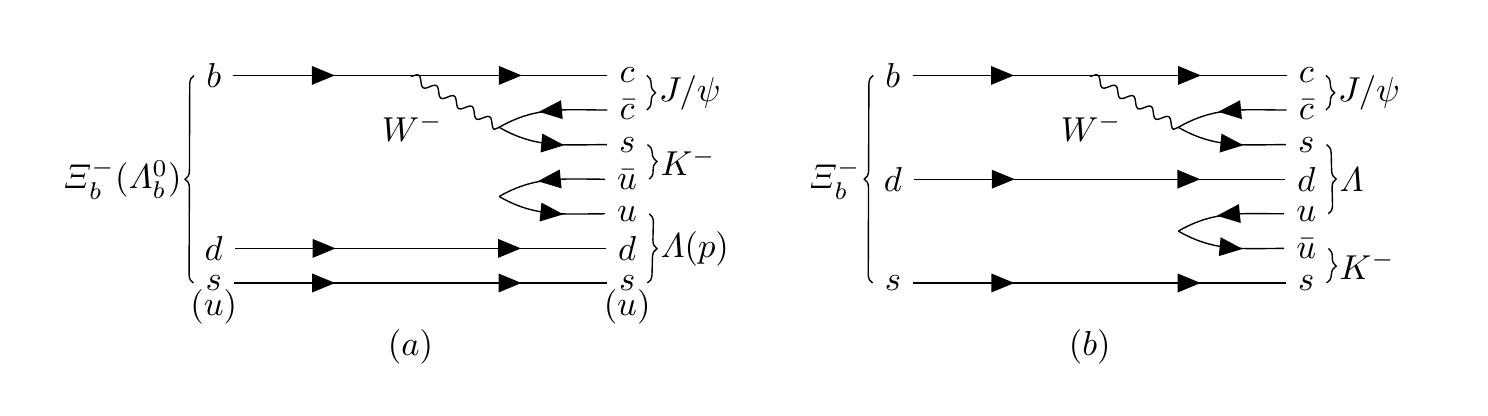}%
\vskip -0.5cm
\caption{Feynman diagrams $\Xibm$ and $\Lb$ decays. Diagram (a) contributes to both \LbJpsipK decays and \XbJpsiLK decays, diagram (b) contributes only to the $\Xibm$ decay.}
\label{fig:feymann1}
\end{figure}

As suggested in Ref.~\cite{Chen:2015sxa}, a hidden-charm pentaquark with open strangeness ($udsc\bar c$)~\cite{Wu:2010jy} could be observed as a $\jpsi\Lz$ state in the decay  \XbJpsiLK. The decay is similar to \LbJpsipK,  and differs from the latter by exchanging one $u$ spectator quark with an $s$ spectator quark, as illustrated in Fig.~\ref{fig:feymann1} (a). An additional diagram can contribute to the $\Xibm$ decay, as illustrated in Fig.~\ref{fig:feymann1} (b), where the $s$ spectator quark forms the $K^-$ meson instead of the $\Lz$ baryon. 

In this Letter, we present the first observation of the $\XbJpsiLK$ decay. Using the decay \LbJpsiL as normalisation channel, the production rate of the observed $\Xibm$ decays relative to that of $\Lb$ baryons is measured as
\begin{equation}
\label{eq:master}
R_{\Xibm/\Lb}\equiv \frac{f_{\Xibm}}{f_{\Lb}}\frac{\BR( \XbJpsiLK)}{\BR( \LbJpsiL)}=\frac{N(\XbJpsiLK)}{N(\LbJpsiL)}\,\epsilon_{\rm rel},
\end{equation}
where $f_{\Xibm}$ and $f_{\Lb}$ are the $b\to\Xibm$ and $b\to\Lb$ fragmentation fractions, $\BR$ represents the branching fraction of the corresponding \bquark-baryon decay, $N(\XbJpsiLK)$ and $N(\LbJpsiL)$ are the signal yields, and $\epsilon_{\rm rel}=\epsilon(\LbJpsiL) / \epsilon(\XbJpsiLK)$ is their relative efficiency. We also present a measurement of the mass difference between the $\Xibm$ and $\Lb$ baryons. Measurements of the $\Xibm$ mass to date have been obtained using absolute mass measurements and a single measurement of the mass difference $\delta M \equiv M(\Xibm)-M(\Lb)$~\cite{PDG2016}. Earlier measurements from the Tevatron~\cite{Aaltonen:2014wfa} are, however, in tension (2.1 standard deviations) with the recent and most precise value from the LHCb experiment~\cite{LHCb-PAPER-2014-048}, obtained from the measurement of $\delta M$. The present analysis offers an opportunity to provide a second precise measurement of $\delta M$ using a data sample that is statistically independent of other measurements of the $\Xibm$ mass from LHCb.

\section{Data sample and detector}
\label{sec:detector}
The measurement is based on a data sample corresponding to 1\invfb of integrated luminosity collected by the \lhcb experiment in $\proton\proton$ collisions at 7\tev centre-of-mass energy in 2011, and 2\invfb at 8\tev in 2012. The \lhcb detector~\cite{Alves:2008zz,LHCb-DP-2014-002} is a single-arm forward
spectrometer covering the \mbox{pseudorapidity} range $2<\eta <5$,
designed for the study of particles containing \bquark or \cquark
quarks. The detector includes a high-precision tracking system
consisting of a silicon-strip vertex detector (\velo) surrounding the $pp$
interaction region, a large-area silicon-strip detector located
upstream of a dipole magnet with a bending power of about
$4{\mathrm{\,Tm}}$, and three stations of silicon-strip detectors and straw
drift tubes placed downstream of the magnet.
The tracking system provides a measurement of momentum, \ptot, of charged particles with
a relative uncertainty that varies from 0.5\% at low momentum to 1.0\% at 200\gevc.
The minimum distance of a track to a primary vertex (PV), the impact parameter (IP), 
is measured with a resolution of $(15+29/\pt)\mum$,
where \pt is the component of the momentum transverse to the beam, in\,\gevc.
Different types of charged hadrons are distinguished using information
from two ring-imaging Cherenkov (RICH) detectors. 
Photons, electrons and hadrons are identified by a calorimeter system consisting of
scintillating-pad and preshower detectors, an electromagnetic
calorimeter and a hadronic calorimeter. Muons are identified by a
system composed of alternating layers of iron and multiwire
proportional chambers.

The online event selection is performed by a trigger~\cite{LHCb-DP-2012-004}, 
which consists of a hardware stage, based on information from the calorimeter and muon
systems, followed by a software stage. For this analysis, triggers that select $\jpsi$ candidates are used for both signal and normalisation channels. The hardware trigger requires at least one muon with $\pt>1.48$ ($1.76$)\gevc, or two muons with $\sqrt{\pt(\mu_1)\pt(\mu_2)}>1.3$ ($1.6$)\gevc, in the 2011 (2012) data sample. 
The subsequent software trigger is composed of two stages, the first of which performs a partial reconstruction and requires either a pair of well-reconstructed, oppositely charged muons having an invariant mass above 2.7\gevcc, or a single well-reconstructed muon with $\pt>1$\gevc and high IP at all PVs of the event. The second stage of the software trigger requires a pair of oppositely charged muons to form a good-quality vertex that is well separated from all PVs, and which has an invariant mass within $\pm120$\mevcc of the known \jpsi mass~\cite{PDG2016}.

In the simulation, $pp$ collisions are generated using
\pythia 8~\cite{Sjostrand:2006za,Sjostrand:2007gs} 
 with a specific \lhcb
configuration~\cite{LHCb-PROC-2010-056}.  Decays of hadronic particles
are described by \evtgen~\cite{Lange:2001uf}, in which final-state
radiation is generated using \photos~\cite{Golonka:2005pn}. The
interaction of the generated particles with the detector, and its response,
are implemented using the \geant
toolkit~\cite{Allison:2006ve, *Agostinelli:2002hh} as described in
Ref.~\cite{LHCb-PROC-2011-006}. The signal decays of \Lb and \Xibm baryons are simulated according to a phase-space model.

\section{Selection requirements}
\label{sec:selection}
The  \XbJpsiLK and  \LbJpsiL candidates are reconstructed using the decays \decay{\jpsi}{\mumu} and \decay{\Lz}{\proton\pim}. An offline selection is applied after the trigger, based on a loose preselection, followed by a multivariate classifier based on a Gradient Boosted Decision Tree (BDTG)~\cite{Breiman}.

In the preselection, the \jpsi candidates are formed from two oppositely charged particles with $\pt>500\mevc$, identified as muons and consistent with originating from a common vertex but inconsistent with originating from any PV. The invariant mass of the $\mumu$ pair is required to be within $[-48, +43]$ \mevcc of the known \jpsi mass~\cite{PDG2016}.

The $\Lz$ candidates are formed by combining candidate $\proton$ and $\pim$ particles with large $\chisqip$, where $\chisqip$ is defined as the difference in the $\chisq$ of the vertex fit for a given PV reconstructed with and without the considered particle. Given the long lifetime of the $\Lz$ baryon, its decay vertex can be reconstructed either from a pair of tracks that include segments in the VELO, called \emph{long} tracks (LL $\Lz$ candidates), or from a pair of tracks reconstructed using only the tracking stations downstream of the VELO, called \emph{downstream} tracks (DD $\Lz$ candidates). 
The invariant mass of the $\proton\pim$ pair is required to be within 4 (6) \mevcc of the known $\Lz$ mass~\cite{PDG2016} for the LL (DD) $\Lz$ candidates. For the LL $\Lz$ candidates, both the proton and the pion must have $\pt>250\mevc$, and pass loose particle identification (PID) criteria based on information provided by the \rich detectors. For the DD $\Lz$ candidates, the decay vertex must not be reconstructed in the first half of the VELO. 
To remove background from \decay{\KS}{\pip\pim} decays, the reconstructed mass for the LL (DD) $\Lz$ candidate under the $\pip\pim$ hypothesis is required to be more than 4 (10) \mevcc away from the known \KS mass~\cite{PDG2016}. 
 
The $\Xibm$ and $\Lb$ candidates are formed from a $\jpsi$ and a $\Lz$ candidate, combined with a kaon candidate for the $\Xibm$ baryon, where the kaon candidate must have $\pt>250$\mevc and large $\chi^{2}_{\rm IP}$. 
Each reconstructed $b$-baryon candidate is required to have $\chi^{2}_{\rm IP}<25$ with respect to at least one PV, and is associated to the one which the $\chi^{2}_{\rm IP}$ is smallest.
The candidate decay vertex must also have a fit with good $\chi^2$ and a separation of at least 1.5\mm from the PV. The angle, $\theta$, between the $b$-baryon momentum and the vector from the associated PV to the decay vertex must satisfy $\cos\theta>0.999$. For both $b$ baryons fiducial cuts of $\pt<25$\gevc and rapidity in the range $2.0<y<4.5$ are required to have a well-defined kinematic region in which the measurement is performed. There are only 0.2\% events outside the fiducial kinematic region. 
A kinematic fit~\cite{Hulsbergen:2005pu} is applied to the \Xibm and \Lb candidates, with the \jpsi and \Lz masses constrained to the known values~\cite{PDG2016}, and the $b$-baryon candidate constrained to point back to its PV. As a result, the mass resolution is improved by 60\%, with most of the improvement coming from the constraints on the $\jpsi$ and \Lz masses. 

The $\XbJpsiLK$ and $\LbJpsiL$ candidates passing the preselection are filtered with a BDTG to further suppress the combinatorial background. For the $\Xibm$ decay, the following discriminating variables are used: the minimum DLL$_{\mu\pion}$ (defined as the difference in the logarithms of the likelihood values from the particle identification systems~\cite{LHCb-DP-2012-003} for the muon and pion hypotheses) and the minimum $\pt$ within the muon pair; the $\chisqip$ of all other final-state tracks and the $\Lz$ baryon; the $\pt$ of the $\proton$, $\pion$, $\kaon$ and $\jpsi$ candidates; the decay length and the vertex fit $\chisq$ of the $\Lz$ candidate; the $\chi^2$ of the kinematic fit, $\cos\theta$ and the decay time of the \Xibm baryon. The BDTG is trained on a simulated \XbJpsiLK sample for the signal; data candidates with $5944<m(\jpsi\Lz\kaon)<6094$\mevcc are used to model the background. 
The LL and DD samples are trained separately. The optimal working point on the BDTG response and the PID variable of the kaon is determined by maximising the significance of the expected $\Xibm$ signal, $S/\sqrt{S+B}$, where $S$ ($B$) is the expected signal (background) yield in a range corresponding to $\pm2.5$ times the mass resolution at the known $\Xibm$ mass~\cite{PDG2016}.  The $S$ value is calculated as the product of an initial signal yield determined from the data at ${\rm BDTG}>0$, and the relative efficiency with respect to the BDTG selection obtained from the simulation. The value of $B$ is estimated from the data sidebands. The final BDTG working point has a signal efficiency of 90\% (70\%) and a background rejection rate of 97\% (99\%) for LL (DD) samples.


The normalisation channel uses a separate training for the BDTG, where the variables for the $K^{-}$ meson are excluded. The background training sample is taken from the $\jpsi\Lz$ invariant mass regions with $150<|m(\jpsi\Lz)-m_{\Lb}|<350$\mevcc, where $m_{\Lb}$ is the known $\Lb$ mass~\cite{PDG2016}. The optimal requirement on the BDTG response for the normalisation mode is the same as for the signal channel. For both samples, in 0.3\% of the cases multiple candidates are found, all of which are retained in the analysis.

\section{Signal yields}
\label{sec:yields}

\begin{figure}[t]
\centering
\vskip -0.3cm
\includegraphics[width=0.5\textwidth]{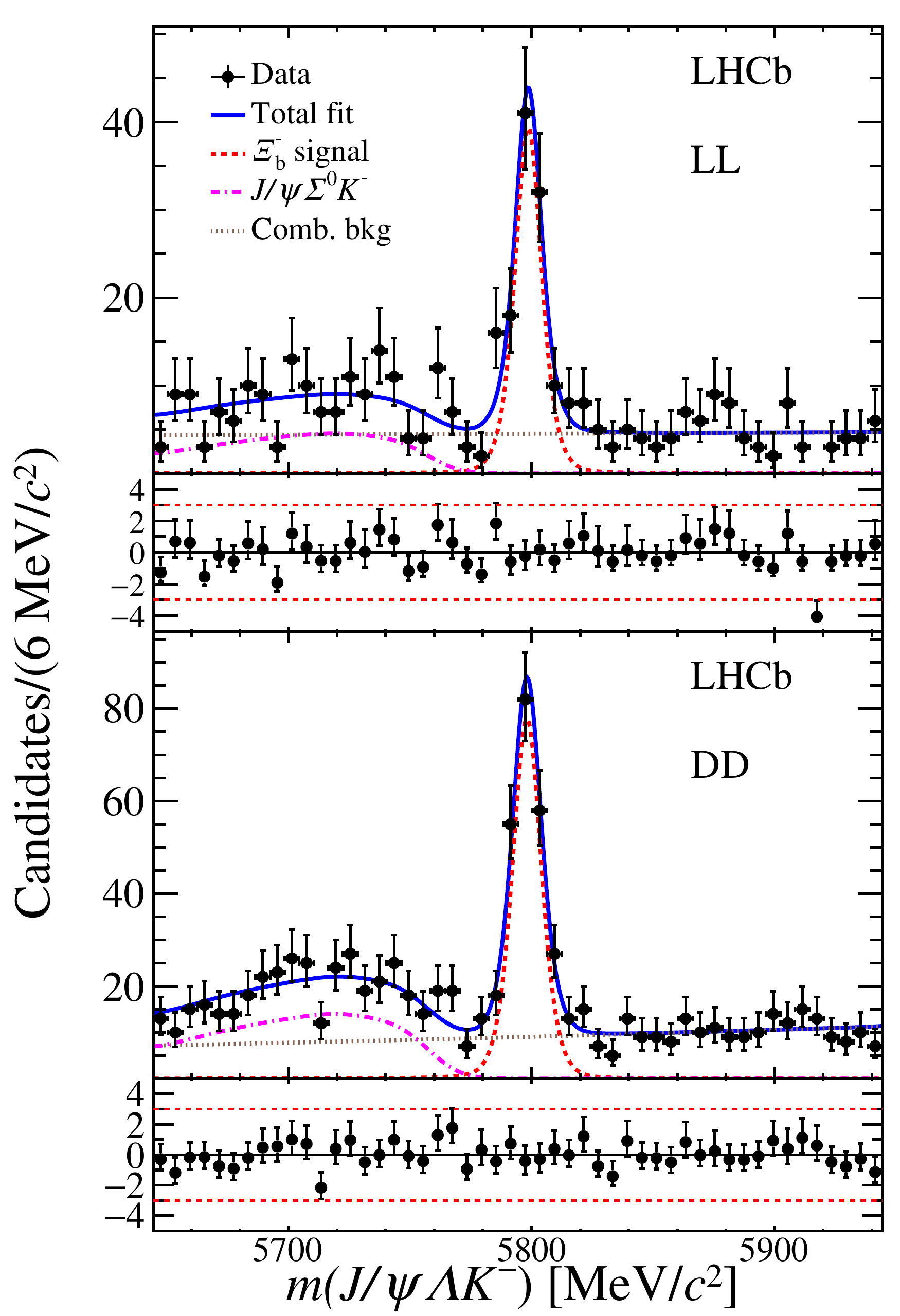}%
\includegraphics[width=0.5\textwidth]{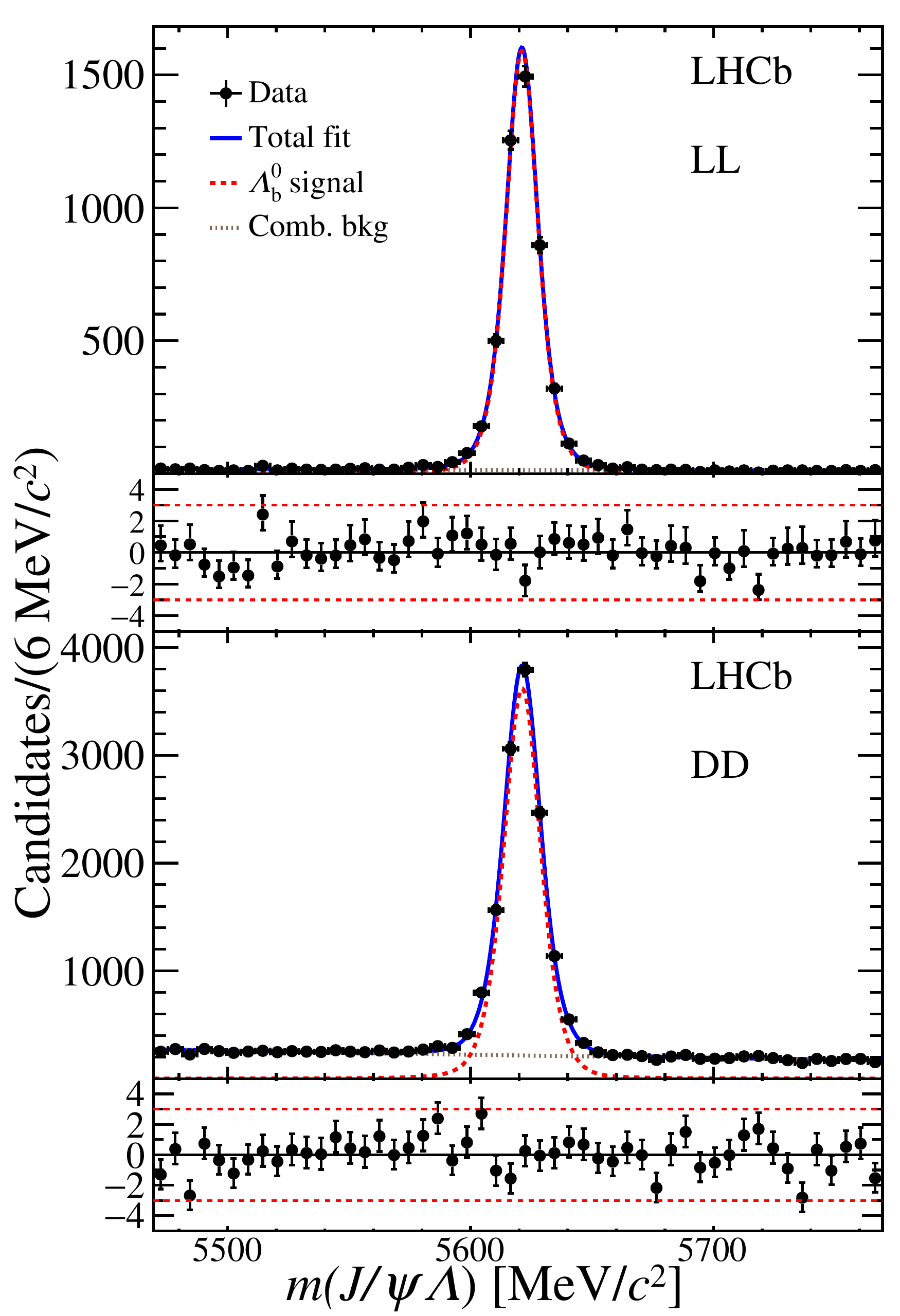}\\
\caption{
Reconstructed (left) \XbJpsiLK and (right) \LbJpsiL candidates using (top) LL and (bottom) DD $\Lz$ types. The solid (blue) lines show the full fit functions, the dashed (red) lines the signal components, the dot-dashed (purple) lines the \XbJpsiSK background and the dotted (black) lines the combinatorial background. At the bottom of each figure the differences between the data and the fit divided by the uncertainty in the data are shown.}
\label{fig:MassFitData}
\vskip -0.3cm
\end{figure}

In each of the two categories (LL and DD), a simultaneous extended unbinned maximum likelihood fit to the $\Xibm$ and $\Lb$ candidates' invariant mass distributions is performed to determine the respective $\Xib$ and $\Lb$ signal yields. The data, separated by category, and the results of the two fits are shown in Fig.~\ref{fig:MassFitData}.

In the fit of each sample, the signal shape is modelled by a Hypatia function~\cite{Santos:2013gra}. 
The mean values and the resolutions of the functions are allowed to vary in the fit, with the ratio of the $\Xibm$ to $\Lb$ mass resolution and the tail parameters fixed to the values obtained from simulation. The combinatorial background is modelled by an exponential function whose parameters are determined by the fit. A partially reconstructed background component, which comes from the decay $\XbJpsiSK$ with $\SLg$, is taken into account in the $\XbJpsiLK$ sample. The shape of this background is determined from simulation, and its yield is free to vary in the fit. In each $\Lz$ category, the fit is simultaneously done for the signal and control channels. The fit procedure is validated by large sets of pseudo-experiments. 

In the LL samples, the signal yields are found to be $N(\XbJpsiLK)=99\pm12$ and $N(\LbJpsiL)=4838\pm72$. The corresponding values in the DD samples are $209\pm17$ and $12\,499\pm125$, respectively. 
The $\XbJpsiSK$ background yields are $72\pm25$ and $221\pm37$ in the LL and DD samples, respectively.
A likelihood-ratio test $\Delta(2\ln{\cal L})\equiv-2\ln(\mathcal{L}_{\rm B}/\mathcal{L}_{\rm S+B})$ is used to estimate the $\XbJpsiLK$ signal significance, where $\mathcal{L}_{\rm B}$ and $\mathcal{L}_{\rm S+B}$ stand for the likelihood values of the background-only hypothesis and the signal plus background hypothesis, respectively. 
A fit to the combined data samples of LL and DD categories is performed to estimate the total signal significance. 
The value of $\Delta(2\ln{\cal L})$ is 464.8. Accounting for two additional parameters associated with the signal component in the $\mathcal{L}_{\rm S+B}$ fit, this corresponds to a significance of 21 standard deviations~\cite{Wilks:1938dza}.

\section{Efficiency corrections}
\label{sec:corrections}

The total efficiency of each decay mode consists of the geometrical acceptance of the detector, the efficiencies of the trigger, the reconstruction and selection, and the hadron identification. The first three efficiency factors are determined from samples of simulated events generated within the kinematic region $\pt<25$\gevc and $2.0<y<4.5$ for both $b$ baryons. The hadron PID efficiency is determined using calibration data of $\Dstarp\to\Dz(\to\Km\pip)\pip$ and $\Lc\to\proton\Km\pip$ decays. Events in the calibration samples are weighted to reproduce the momentum, pseudorapidity and event multiplicity distributions of the hadrons from $\XbJpsiLK$ and $\LbJpsiL$ decays. The relative efficiency is estimated to be $\epsilon_{\rm rel}=\epsilon(\LbJpsiL) / \epsilon(\XbJpsiLK)=1.964\pm0.028$ and $2.191\pm 0.017$ for the LL and DD samples, respectively, where the uncertainties are statistical only.

Two correction factors are considered for the relative efficiency to account for differences between data and simulation. The LL and DD samples are combined to derive these factors. The first factor accounts for possible local structures in the data distribution due to intermediate states or nonresonant amplitudes that are generally present in multibody decays. An average efficiency is calculated over the two-dimensional phase space of the $\XbJpsiLK$ three-body decay,
\begin{equation}\label{eq:eff}
\langle\epsilon\rangle = \sum_i w_i  / \sum_i (w_i/\epsilon_{{\rm PH}\,i}),
\end{equation}
where $\epsilon_{\rm PH}$ is the efficiency as a function of the phase-space position obtained from simulation, the numerator represents the number of reconstructed signal candidates, and the denominator represents the efficiency-corrected number of signal candidates; in both cases the sum extends over all $\Xibm$ candidates in data. The event-by-event signal weight ($w_i$), is obtained using the \sPlot technique~\cite{Pivk:2004ty} to subtract the background contribution. The average efficiency is 98\% relative to the efficiency obtained using the phase-space simulation.

The second factor accounts for possible differences in $\pt$ and rapidity spectra in $b$-baryon production in data and simulation.  The simulated samples are reweighted in bins of $\pt$ and rapidity, in order to reproduce the data distribution of $\Lb$ decays, and the relative efficiency is recalculated. The correction factor of this source is $1.138$. The value is consistent if separately correcting for the LL and DD samples. The product of the two correction factors for the average efficiency is $1.115$. The uncertainties in the correction factors are taken as systematic uncertainties discussed below.

\section{\boldmath Results of $R_{\Xibm/\Lb}$ and systematic uncertainties}
\label{sec:systematics}
Using the yields and efficiencies with corrections, the ratios of $R_{\Xibm/\Lb}$ for the LL and DD data sets are measured to be $(4.46\pm 0.55)\times10^{-2}$ and $(4.08\pm0.34)\times10^{-2}$,
respectively, where the uncertainties are statistical only. The two independent measurements are consistent with each other. Their weighted average yields 
$R_{\Xibm/\Lb}=(4.19\pm 0.29\pm0.15)\times10^{-2}$.
Whenever two uncertainties are quoted, the first is statistical and the second is systematic.

\begin{table}[b] 
\centering
\caption{Relative systematic uncertainty for the ratio $R_{\Xibm/\Lb}$.}
\label{tab:systemerr}
\begin{tabular}{lc}
Source&Uncertainty (\%)\\\hline
Signal model&0.7\\
Background model &1.6\\
BDTG efficiency& 0.1\\
PID efficiency&1.0\\
Tracking efficiency& 1.2\\
Phase space& 1.5\\
$b$-baryon kinematics & 1.5\\
\Xibm and \Lb lifetime &  1.1\\
Simulation sample size & 0.7\\
Fixed resolution ratio & 0.6\\
\hline
Total& 3.5\\
\end{tabular}
\end{table}

The sources of systematic uncertainties for the ratio $R_{\Xibm/\Lb}$ are summarised in Table~\ref{tab:systemerr}. The quoted values are averages over the LL and DD categories. 
The uncertainty on the relative yields is evaluated by using alternative functions to model each of the fit components. 
These include 
changing the signal model from the Hypatia function to a double-sided Crystal Ball function~\cite{Skwarnicki:1986xj}, changing the combinatorial background model from the exponential function to a second-order polynomial, and varying
the parametrisation of the \XbJpsiSK background. The effect of the latter is found to be negligible. To reduce the statistical fluctuations in the estimate of the systematic uncertainties, large numbers of pseudoexperiments are performed. The parameters of the alternative model are used to generate experiments, which are then fitted by both the alternative and the default models. 
A Gaussian function is fitted to the distribution of the $R_{\Xibm/\Lb}$ difference for these pseudoexperiments and the mean value is assigned as a systematic uncertainty.

There are several sources of systematic uncertainty related to the evaluation of the relative efficiency. Most of them cancel in the ratio of efficiencies, except those related to the additional kaon in the $\Xibm$ decay. The BDTG input variables for background-subtracted $\LbJpsiL$ data are compared to the corresponding simulated distributions, and all of the variables, except for the vertex-fit $\chi^{2}$ and $\chisqip$ for $\Lz$ candidates in the DD category, are well modelled. The simulation is then smeared for these two variables to match the data, and the small change of 0.1\% in the relative efficiency is taken as systematic uncertainty.
The uncertainty due to the kaon PID efficiency is studied by changing the binning scheme in momentum, pseudorapidity and event multiplicity. The alternative binning gives a 1.0\% difference in the signal efficiency, which is assigned as a systematic uncertainty.
The tracking efficiency is estimated from simulation and calibrated with the data~\cite{LHCb-DP-2013-002}; an uncertainty of 0.4\% is assigned for the kaon track. 
An additional systematic uncertainty of 1.1\% is assigned to the kaon tracking efficiency due to an imperfect knowledge of the material budget in the detector~\cite{LHCb-PAPER-2015-032}. It is estimated from simulation by changing the used interaction length in the detector by 10\%. 
The total tracking-efficiency related systematic uncertainty, adding the two contributions in quadrature, is 1.2\%.

The systematic uncertainty of the average efficiency defined in Eq.~(\ref{eq:eff}) is 1.5\%, calculated by propagating the statistical uncertainties for the efficiencies over the phase space. In reweighting the simulated $\pt$ and $y$ spectra to match the data, an uncertainty of 1.5\% is estimated by varying the weights for each kinematic bin by its uncertainty.  The uncertainties in the \Lb lifetime of $1.468\pm0.012$\ps~\cite{LHCb-PAPER-2014-003} and the $\Xibm$ lifetime of $1.57\pm0.04$\ps~\cite{PDG2016}, result in relative changes of $\pm0.2\%$ and $\pm1.1\%$ in the efficiencies, respectively.
The limited size of the simulated samples gives rise to an uncertainty of 0.7\%. 
Varying the mass resolution ratios of the $\Xib$ to $\Lb$ mass peaks, which are fixed in the nominal fit to the data, results in an uncertainty of 0.6\%.
The uncertainty due to the trigger efficiency is cancelled between the signal and control modes, as the trigger requirements are imposed only on the muon pairs. Finally, the total relative systematic uncertainty is 3.5\%, obtained by adding all of the above contributions in quadrature.

\section{Measurement of the mass difference}
\label{sec:results}

The mass difference, $\delta M$, is obtained from a single simultaneous fit to four mass distributions, consisting of the LL and DD samples for both the $\Xibm$ and $\Lb$ candidates.
The ratio $R_{\Xibm/\Lb}$ is also a freely varying parameter in this second fit for $\delta M$.
Compared to the fits described in the previous section, the new fit has two less free parameters: for each of the $\Lz$ categories, $\delta M$ is constrained to be the same value and $N(\XbJpsiLK)$ is replaced by $N(\LbJpsiL)*\epsilon_{\rm rel}*R_{\Xibm/\Lb}$.
The simultaneous fit gives the same result as the weighted average for the ratio $R_{\Xibm/\Lb}$, and the mass difference is measured to be 
\begin{equation}
\delta M  = 177.08\pm0.47\pm0.16\mevcc. \nonumber
\end{equation}
This measurement is of similar precision to and consistent with the previous \lhcb result $\delta M=178.36\pm0.46\pm0.16$\mevcc using $\Xibm\to\Xicz \pi^-$ and $\Lb\to \Lc\pi^-$ decays~\cite{LHCb-PAPER-2014-048}. 
The two results are combined to obtain $\delta M=177.73\pm0.33\pm0.14$\mevcc, where the correlations between the systematic uncertainties described below are properly taken into account.

Various sources of systematic uncertainty are considered for the mass difference measurement. The effect of the momentum scale uncertainty of 0.03\%~\cite{LHCb-PAPER-2013-011} leads to an uncertainty of 0.13\mevcc. Because the signal mode has one more particle than the normalisation channel, the correction for energy loss in the detector material leads to an additional uncertainty of 0.06\mevcc~\cite{LHCb-PAPER-2013-011,LHCb-PAPER-2014-048}. The above two sources are fully correlated with the previous measurement using $\Xibm\to\Xicz \pi^-$ and $\Lb\to \Lc\pi^-$ decays~\cite{LHCb-PAPER-2014-048}. Uncertainties due to the signal and background modelling are 0.06 and 0.02\mevcc, respectively, estimated by considering alternative functions as discussed in Sec.~\ref{sec:systematics}.

\section{Conclusion}
In conclusion, we report the first observation of the \XbJpsiLK decay with a data sample of $\proton\proton$ collisions corresponding to an integrated luminosity of 3\invfb.  The observed signal yield is $308\pm21$. 
In the kinematic region of the $b$-baryon transverse momentum $\pt<25$\gevc and rapidity in the range $2.0<y<4.5$, the production rate of $\Xibm$ with $\XbJpsiLK$ decays relative to that of $\LbJpsiL$ decays is measured to be
\begin{equation*}
\frac{f_{\Xibm}}{f_{\Lb}}\frac{\BR(\XbJpsiLK)}{\BR(\LbJpsiL)}=(4.19\pm 0.29\stat \pm0.15 \syst)\times 10^{-2},
\end{equation*}
where $f_{\Xibm}/f_{\Lb}$ is the ratio of the fragmentation fraction for $b\to\Xibm$ and $b\to\Lb$ transitions. The mass difference between $\Xibm$ and $\Lb$ baryons is measured to be 
\begin{equation*}
M(\Xibm)-M(\Lb)=177.08\pm0.47\stat\pm0.16\syst \mevcc. 
\end{equation*}
A combination of this value with the previous LHCb measurement from $\Xibm\to\Xicz \pi^-$ and $\Lb\to \Lc\pi^-$ decays~\cite{LHCb-PAPER-2014-048} leads to the most precise value of the mass difference 
\begin{equation*}
M(\Xibm)-M(\Lb)=177.73\pm0.33\stat\pm0.14\syst \mevcc. 
\end{equation*}

With the full data sample accumulated before the long shutdown of the LHC in 2018, it should be possible to apply a full amplitude analysis to the $\XbJpsiLK$ decay to search for hidden-charm pentaquarks with open strangeness.

\section*{Acknowledgements}
\noindent We express our gratitude to our colleagues in the CERN
accelerator departments for the excellent performance of the LHC. We
thank the technical and administrative staff at the LHCb
institutes. We acknowledge support from CERN and from the national
agencies: CAPES, CNPq, FAPERJ and FINEP (Brazil); NSFC (China);
CNRS/IN2P3 (France); BMBF, DFG and MPG (Germany); INFN (Italy); 
FOM and NWO (The Netherlands); MNiSW and NCN (Poland); MEN/IFA (Romania); 
MinES and FASO (Russia); MinECo (Spain); SNSF and SER (Switzerland); 
NASU (Ukraine); STFC (United Kingdom); NSF (USA).
We acknowledge the computing resources that are provided by CERN, IN2P3 (France), KIT and DESY (Germany), INFN (Italy), SURF (The Netherlands), PIC (Spain), GridPP (United Kingdom), RRCKI and Yandex LLC (Russia), CSCS (Switzerland), IFIN-HH (Romania), CBPF (Brazil), PL-GRID (Poland) and OSC (USA). We are indebted to the communities behind the multiple open 
source software packages on which we depend.
Individual groups or members have received support from AvH Foundation (Germany),
EPLANET, Marie Sk\l{}odowska-Curie Actions and ERC (European Union), 
Conseil G\'{e}n\'{e}ral de Haute-Savoie, Labex ENIGMASS and OCEVU, 
R\'{e}gion Auvergne (France), RFBR and Yandex LLC (Russia), GVA, XuntaGal and GENCAT (Spain), Herchel Smith Fund, The Royal Society, Royal Commission for the Exhibition of 1851 and the Leverhulme Trust (United Kingdom).

\addcontentsline{toc}{section}{References}
\setboolean{inbibliography}{true}
\bibliographystyle{LHCb}
\bibliography{main,LHCb-PAPER,LHCb-CONF,LHCb-DP,LHCb-TDR,MyBib}

\newpage
 
\newpage
\centerline{\large\bf LHCb collaboration}
\begin{flushleft}
\small
R.~Aaij$^{40}$,
B.~Adeva$^{39}$,
M.~Adinolfi$^{48}$,
Z.~Ajaltouni$^{5}$,
S.~Akar$^{59}$,
J.~Albrecht$^{10}$,
F.~Alessio$^{40}$,
M.~Alexander$^{53}$,
S.~Ali$^{43}$,
G.~Alkhazov$^{31}$
% ...

{\footnotesize \it
% ...
$ ^{5}$Clermont Universit{\'e}, Universit{\'e} Blaise Pascal, CNRS/IN2P3, LPC, Clermont-Ferrand, France\\
% ...
$ ^{10}$Fakult{\"a}t Physik, Technische Universit{\"a}t Dortmund, Dortmund, Germany\\
% ...
$ ^{31}$Petersburg Nuclear Physics Institute (PNPI), Gatchina, Russia\\
% ...
$ ^{39}$Universidad de Santiago de Compostela, Santiago de Compostela, Spain\\
$ ^{40}$European Organization for Nuclear Research (CERN), Geneva, Switzerland\\
% ...
$ ^{43}$Nikhef National Institute for Subatomic Physics, Amsterdam, The Netherlands\\
% ...
$ ^{48}$H.H. Wills Physics Laboratory, University of Bristol, Bristol, United Kingdom\\
% ...
$ ^{53}$School of Physics and Astronomy, University of Glasgow, Glasgow, United Kingdom\\
% ...
$ ^{59}$University of Cincinnati, Cincinnati, OH, United States\\
% ...
}
\end{flushleft}

\end{document}